\documentclass[twocolumn,showpacs,aps]{revtex4}
\usepackage{amssymb}
\usepackage{epsfig}
\usepackage{amsfonts}
\usepackage{amsmath}
\usepackage{amssymb}
\usepackage{graphicx}

\begin{document}

\title[EM Solitons in Degenerate E-P Plasma]{Electromagnetic Solitons in Degenerate
Relativistic Electron-Positron Plasma}
\author{V.I. Berezhiani$^{1,2}$, N.L. Shatashvili$^{1,3}$ and N.L. Tsintsadze$^{1,3}$}
\address{
$^{1}$Andronikashvili Institute of Physics, TSU, Tbilisi \ 0177,
Georgia \\
$^{2}$School of Physics, Free University of Tbilisi, Georgia \\
$^{3}$Department of Physics, Faculty of Exact and Natural
Sciences, Ivane Javakhishvili Tbilisi State University, TSU,
Tbilisi 0179, Georgia }


\begin{abstract}
The existence of soliton-like electromagnetic (EM) distributions
in a fully degenerate electron-positron plasma is studied applying
relativistic hydrodynamic and Maxwell equations. For circularly
polarized wave it is found that the soliton solutions exist both
in relativistic as well as nonrelativistic degenerate plasmas.
Plasma density in the region of soliton pulse localization is
reduced considerably. The possibility of plasma cavitation is also
shown.
\end{abstract}


\maketitle

\section{Introduction}

During the past few years considerable amount of papers have been
devoted to the analysis of electromagnetic (EM) wave dynamics in
relativistic plasmas primarily in connection with their possible
role in a variety of astrophysical phenomena. Highly relativistic
electron-positron (e-p) plasmas exist in the pulsar magnetosphere
\cite{bib:Sturok} and the corona of the magnetars
\cite{bib:Beloborodov}, and also likely to be found in the bipolar
outflows (jets) in Active Galactic Nuclei (AGN)
\cite{bib:Begelman}. The presence of e-p plasma is also argued in
the MeV epoch of the early Universe \cite{bib:Tajima}. The plasma
can be termed as a relativistic when either bulk velocities of
plasma fluid cells are close to the velocity of light or when the
averaged kinetic energy of the particles in the cells are greater
than the electron rest energy (e.g. the thermal energy of plasma
with temperature $T\geq mc^{2}$). In different astrophysical
conditions the density of e-p plasmas can take values varying by
many orders of magnitude. It is believed that the rest frame
density of the e-p plasma near the pulsar surface is $n\geq
10^{11}\,cm^{-3}$ \cite{bib:Misha}, while in the MeV epoch of the
early Universe the density of the optically thick e-p plasma can
be as high as $n=10^{32}\,cm^{-3}$ \cite{bib:Weinberg}. Intense
e-p pair creation takes place during the process of gravitational
collapse of massive stars \cite{bib:Stenflo}. It is argued that
the gravitational collapse of the massive stars may lead to the
charge separation with the field strength exceeding the Schwinger
limit resulting in e-p pair plasma creation with estimated density
to be $n=10^{34}\,cm^{-3}$ \ \cite{bib:ruffini}. Superdense e-p
plasma may also exist in GRB source where the e-p plasma density
can be in the range $n=(10^{30}-10^{37})\,cm^{-3}$
\cite{bib:aksenov}. Dense electron-positron plasma can be soon
produced in laboratory conditions as well. Indeed, the modern
petawatt lasers systems are already capable of producing
ultrashort pulses with the focal intensities \ $I=2\times
10^{22}\,W/cm^{2}$ \ \cite{bib:Yanovski}. Pulses of even higher
intensities exceeding \ $I=10^{26}\,W/cm^{2}$ \ are likely to be
available soon in lab or in the Lorentz boosted frames
\cite{bib:Dunne}. Interaction of such pulses with gaseous or solid
targets could lead to the generation of the optically thin e-p
plasma with above solid state densities in the range of \
$(10^{23}-10^{28})\,cm^{-3}$ \ \cite{bib:Shukla-Eliasson}.

For highly compressed state the plasma behaves as a degenerate
Fermi gas provided that averaged inter-particle distance is
smaller than the thermal de Broglie wavelength. Mutual interaction
of the plasma particles becomes unimportant and plasma becomes
more ideal as the density increases \cite{bib:Landau}. If the
thermal energy of the particles (electrons and positrons) is much
lower than their Fermi energy the plasma may be treated as cold,
i.e. having zero temperature, even it is of the order of \
$10^{9}\,K$ \ \cite{bib:Russo}. The Fermi energy of degenerate
electrons (positrons) is \ \ $\epsilon _{F}=m_{e}c^{2}\,\left[
\left( 1+R^{2}\right) ^{1/2}-1\right] $ , where \ $R=p_{F}/m_{e}c$
, \ $p_{F}$ \ - is the Fermi momentum which is related to the
rest-frame particle density by the following relation \
$p_{F}=m_{e}c\,\left( n/n_{c}\right) ^{1/3}$ \ , \ here \
$n_{c}=5.9\times 10^{29}\,cm^{-3}$ \ is the normalizing critical
number-density \cite{bib:Akbari}. Thus, if  \ $n\geq n_{c}$ \ then
particles inside fluid cells can move with the relativistic
velocities and plasma rightfully can be termed as being a
relativistic. Here we would like to emphasize that the pair
plasmas with such densities can not be in complete thermodynamic
equilibrium with the photon gas into which it annihilates
\cite{bib:Katz}. Equilibrium is reached within the time-period
related mainly to the electron-positron annihilations. The
characteristic time for the e-p pair annihilation is \
$\tau_{ann}\approx 1/(\sigma v\,n)$ \cite{bib:zeldovich} where
$\sigma $ is the cross section of annihilation and $v$ is the
relative velocity of pairs. In relativistic case, for the plasma
density \ $n=10^{30}\,cm^{-3}$ \ (i.e. $p_F=m_ec$) \ the particles
move with almost velocity of light \ $v\approx c$ \ and for
annihilation cross section roughly estimated to be \ $\sigma
\approx 10^{-24}cm^{2}$ \ we get \ $\tau_{ann}\approx 0.3\times
10^{-16}\,sec$. Subsequently, the thermodynamic equilibrium
between pairs and photons (with zero chemical potential) will be
achieved. Plasma becomes optically thick with steady state pair
density defined by plasma temperature. For low temperature
degenerate plasma the steady state pair density will be
considerably smaller than the initial one. Therefore, in order to
study electromagnetic properties of relativistic degenerate e-p
plasma the characteristic time of the plasma processes should be
smaller than the time of pair annihilations. Although the pair
annihilation time turns out to be extremely small the
characteristic plasma frequency (Langmuir frequency) for high
densities is also very high \ [$\sim 10^{18}\,sec^{-1}$] \ and
collective plasma oscillations have enough time to develop.

\section{Basic Equations}

In the present paper we study the possibility of the existence of
localized high frequency EM solitary pulses in the relativistic
degenerate optically thin e-p plasma. Existence and stability of
EM solitary pulses in classical relativistic e-p plasma has been
intensively investigated in the past
\cite{bib:physreport,bib:levan,bib:kartal,bib:lee}). In contrast,
the nonlinear dynamics of EM pulses in quantum degenerate
relativistic plasma were investigated just for low frequency modes
(see \cite{bib:Khan} and references therein).

The fluid equations, valid for each species (electrons and
positrons), can be written in a manifestly covariant form
\begin{equation}
\frac{\partial T^{\alpha \beta }}{\partial x^{\beta }} =
qF^{\alpha \beta }nU_{\beta }\ ,
\label{B1}
\end{equation}
where the Greek indices go from \ $0$ \ to \ $3$ ; here \
$\partial _{\alpha }\equiv \partial /\partial x^{\alpha }=\left(
c^{-1}\partial /\partial t, \mathbf{\nabla }\right) $; \
$T^{\alpha \beta }$ \ is the energy-momentum tensor of the plasma
species with charge \ $q$ \ and mass \ $m$ , and \ $U^{\alpha
}=\left( \gamma ,\gamma \mathbf{V/}c\right) $ \ is the local
four-velocity with \ $\gamma =\left( 1-V^{2}/c^{2}\right) ^{-1/2}$
$\left( U^{\alpha }U_{\alpha }=1\right) $ ; the metric tensor is \
$g^{\alpha \beta }=diag\left( 1,-1,-1,-1\right) $ , \ and \
$q=-e,\ e$ \ for electrons and positrons respectively. Note that
we do not label the fluid species by an additional index for
brevity.

In case when the number of charged particles is conserved for each
species, the rest-frame particle density \ $n$ \ satisfies the
continuity equation
\begin{equation}
\frac{\partial nU^{\alpha}}{\partial x^{\alpha}}=0 \ .
\label{B2}
\end{equation}

\bigskip

The electromagnetic (EM) field tensor can be formally written as \
$F^{\alpha\beta}=[{\bf E},{\bf B}]$ \ and it satisfies the Maxwell
equations \ $\partial_{\beta}F^{\alpha\beta}=-(4\pi/c)J^{\alpha}$
, \qquad
$\epsilon^{\alpha\beta\gamma\delta}\partial_{\beta}F_{\gamma\delta}=0$,
\ where \ $J^{\alpha}=(c\rho,{\bf J}$ $)$ , and \ $\rho$ \ and \
$\mathbf{J} $ \ are the total charge and the current density of
the plasma, respectively. Equation (\ref{B1}) represents the
conservation of momentum and energy, where the momentum change due
to collisions is ignored.

The energy momentum tensor \ $T^{\alpha\beta}$ \ is assumed to be
that of an ideal isotropic fluid: \
$T^{\alpha\beta}=wU^{\alpha}U^{\beta}-g^{\alpha
\beta}{\mathcal{P}}$ , where \ $w={\mathcal{E}}+{\mathcal{P}}$ \
is the enthalpy per unit volume, and \ ${\mathcal{E}}$ \ is the
proper internal energy density, while \ ${\mathcal{P}}$ \ is the
pressure of the fluid species. If the thermal energy of the
particles (electrons and positrons) is much lower than their Fermi
energy the plasma can be treated as being completely degenerate.
For a degenerate Fermi gas with \ $nT/{\mathcal{P}}<<1 $ , we have
the following relations \cite{bib:Chandra,bib:gurovich} :
\begin{equation}
{\mathcal{P}} = \frac{m_{e}^{4}c^{5}}{3\pi^{2}\hbar^{3}}f(R) \ ,
\label{B3}
\end{equation}

\begin{equation}
{\mathcal{E}} =\frac{m_{e}^{4}c^{5}}{3\pi^{2}\hbar^{3}}\left[ R^{3}\left(
1+R^{2}\right) ^{1/2}-f(R) \right] \ ,
\label{B4}
\end{equation}
where
\begin{equation}
8f\left( R\right) =3\sinh^{-1}R+R\left( 1+R^{2}\right) ^{1/2}\left(
2R^{2}-3\right) \ .
\label{B5}
\end{equation}

In equations (\ref{B3})-(\ref{B5}) \ $R=p_{F}/m_{e}c$ \ with \ $p_{F}$ \
being the Fermi momentum defined above. The degenerate equation of state is
given by \ ${\mathcal{P}}\propto$ $n^{5/3}$ \ or \ ${\mathcal{P}}\propto
n^{4/3}$ \ for nonrelativistic \ $\left( R\ll 1\right)$ \ and
ultrarelativistic \ $\left( R\gg 1\right)$ \ plasma cases, respectively.

\bigskip

The fluid model presented above implies that the distribution
function of electrons and positrons remains locally
Juttner-Fermian which for zero temperature case leads to the just
density dependent thermodynamical quantities \ ${\mathcal{E}}(n),
\ {\mathcal{P}}(n)$ \ and \ $w(n)$ . All these quantities
implicitly depend on \ $x_{\alpha}$ \ via \ $n=N/\gamma$ , where \
$N$ \ is the density of the fluid species in laboratory frame. The
plasma dynamics is isentropic (moreover, at \ $T\rightarrow 0$ \
the entropy turns out to be zero) and, consequently, we have the
following thermodynamical relation \ $d( w/n ) =d{\mathcal{P}}/n$.
Applying this relation and passing through straightforward algebra
(see for instance \cite{bib:Ohashi,bib:BMY-cooling}), introducing
\ $G=w/nm_{e}c^{2}$ \ , Eq. (\ref{B1}) can be reduced to the
following system of equations:
\begin{equation}
\frac{\partial}{\partial t}\left(G{\bf p}\right)+ m_{e}c^{2}\,
{\bf \nabla}\left(G\gamma\right) = q\,{\bf E} \ + \ {\bf V\times
\Omega}
\label{B6}
\end{equation}
and
\begin{equation}
\frac{\partial}{\partial t}\,{\bf \Omega}={\bf \nabla\times}\left(
{\bf V\times\Omega}\right) \ ,
\label{B7}
\end{equation}
where \ ${\bf \Omega=}( q/c ) {\bf B+\nabla\times}\left( G{\bf
p}\right)$ \ is the generalized vorticity. Here \ ${\bf p} =
\gamma m_{e}{\bf V}$ \ is the hydrodynamic momentum and \ $G=G(n)$
\ can be called as the density dependent "effective mass" factor
of the fluid cell. \ The Equations (\ref{B6})-(\ref{B7}) along
with the Maxwell and the continuity equations form the complete
set of equations for studying the dynamics of degenerate plasma.
It is interesting to remark that similar set of equations has been
exhibited in Ref.\cite{bib:Ohashi} for classical relativistic
plasma obeying Maxwell-Juttner statistics. In this case the
effective mass factor of the fluid elements depends on the
temperature \ $G=G(T)$ \ whereas \ $T\sim n^{\Gamma-1}$. \ Here
the adiabatic index \ $\Gamma=5/3$ \ for the nonrelativistic \
($T\ll m_{e}c^{2}$) \ and $\Gamma=4/3 $ \ for the relativistic \
($T\gg m_{e}c^{2}$) \ temperatures, respectively. In the
degenerate plasma case \ $w/nm_{e}c^{2}=\left( 1+R^{2}\right)
^{1/2}$ \ and consequently the mass factor depends only on the
plasma rest frame density by the following simple relation \ $G=[
1+( n/n_{c})^{2/3}]^{1/2}$ \ which is valid for the arbitrary
strength of relativity defined by the ratio \ - $n/n_c$ .

\vspace{1cm}

Expressing the EM fields by the vector \ (${\bf A}$) \ and scalar
\ ($\varphi $) \ potentials, i.e., \ ${\bf E} = -(1/c){\bf
\partial A/\partial}t-{\bf \nabla}\varphi$ , \ and \
${\bf B}={\bf \nabla\times A}$ , the Maxwell equations (with the
Coulomb gauge \ ${\bf \nabla\cdot A}=0$ ) \ can be written as:

\begin{equation}
\frac{\partial^{2}{\bf A}}{\partial t^{2}}-c^{2}\Delta{\bf A} +
c\, \frac{\partial}{\partial t}\left( {\bf \nabla}\varphi\right) -
4\pi c\,{\bf J}=0
\label{B8}
\end{equation}
and

\begin{equation}
\Delta\varphi = - 4\pi\,\rho \ .
\label{B9}
\end{equation}

\noindent Here, the charge and the current densities are given by
\ $\rho=\sum q\gamma n$ \ and \ ${\bf J}=\sum q\gamma n{\bf V}$,
respectively; the summation runs over the charge species. For the
current effort, we apply equations (\ref{B6})-(\ref{B7}) for wave
processes in an unmagnetized plasma. From Eq.(\ref{B7}) it follows
that if the generalized vorticity is initially zero \ (${\bf
\Omega} = 0$) \ everywhere in space, it remains zero for all
subsequent times. We assume that before the EM radiation is
''switched on'' the generalized vorticity of the system is zero.
Accordingly, the Eq.(\ref{B6}) now takes the form

\begin{equation}
\frac{\partial}{\partial t}\left( G{\bf p} + \frac{q}{c}{\bf
A}\right) +{\bf \nabla}\left( m_{e}c^{2}\,G\,\gamma +
q\,\varphi\right) = 0 \ . \label{B10}
\end{equation}

\bigskip

We are looking for the localized solution of equations
(\ref{B8})-(\ref{B10}) in one-dimensional case. Assuming that all
quantities vary only with one spatial coordinate \ $z$ \ and in
time \ $t$ \ the transverse component of the equation of motion
(\ref{B10}) is immediately integrated to give: \ ${\bf p}_{\perp}
= - q\,{\bf A}_{\perp}/(c\,G)$ . The constant of integration is
set equal to zero, since the particle hydrodynamic momenta are
assumed to be zero at infinity where the field vanishes. \ Due to
the gauge condition \ $A_{z}=0$ \ the longitudinal motion of the
plasma is coupled with the EM field via \ $\gamma=\left[1+(
p_{\perp}^{2}+p_{z}^{2})/m_{e}^{2}c^{2}\right]^{1/2}$ \ which does
not depend on the particle charge sign. The EM pressure gives
equal longitudinal momenta to both the electrons and positrons \
($p_{ez} = p_{pz}=p_{z}, \ \gamma_{e}=\gamma_{p}=\gamma$) \ and
modifies plasma density without producing the charge separation,
i.e., \ $n_{e}=n_{p}=n$ \ and \ $\varphi =0$ . \ Thus, the
longitudinal motion of the plasma is entirely determined by the
following equation of motion
\begin{equation}
\frac{\partial}{\partial
t}G\,p_{z}+m_{e}c^{2}\,\frac{\partial}{\partial z}\,G\,\gamma=0 \
\label{B11}
\end{equation}
and the continuity equation (\ref{B2}), which in our case reads as
\begin{equation}
\frac{\partial}{\partial t}\gamma n+\frac{\partial}{\partial z}\left(
n\gamma V_{z}\right) = 0 \ .
\label{B12}
\end{equation}

\noindent The longitudinal component of the current density is
zero \ $J_{z}=0$ \ while for the transverse one we have: \ ${\bf
J}_{_{\perp}}{\bf =}( 2ne^2/c\,G ) {\bf A}_{\perp}$ . Substituting
this expression into Eq.(\ref{B8}) we obtain the following
equation

\begin{equation}
\frac{\partial^{2}{\bf A}_{\perp}}{\partial
t^{2}}-c^{2}\frac{\partial ^{2}{\bf A}_{\perp}}{\partial z^{2}} +
\Omega_{e}^{2}\left( \frac{n}{n_{0}}\frac{G_{0}}{G}\right) {\bf
A}_{\perp}=0 \ ,
\label{B13}
\end{equation}
where \ $\Omega_{e}=( 8\pi e^{2}n_0/(m_{e}G_{0})\ )^{1/2}\ $ \ is
the Langmuir frequency of the e-p plasma and \ $n_{0}$ \ is the
equilibrium density of electrons (positrons). Notice that in the
Langmuir frequency we introduced effective electron mass \
$m_{e}G_{0}=m_{e}(1+R_{0}^{2})^{1/2}$ , where \
$R_{0}=(n_{0}/n_{c})^{1/3}$.

\bigskip

We are looking for the stationary localized solution described by
the equations (\ref{B11})-(\ref{B13}) for a circularly polarized
EM wave. The vector potential can be expressed as
\begin{equation}
e{\bf A}_{\perp}/m_{e}c^{2} =(1/2)({\bf x} + i{\bf
y})\,A(z)\exp(-i\omega t)+c.c. \ ,
\end{equation}
where \ $A(z)$ \ is the real valued dimensionless amplitude
depending only on spatial coordinate \ $z$; \ $\omega$ \ is the
frequency and \ ${\bf x}$ \ and \ ${\bf y}$ \ are the unit
vectors. In this stationary case \ $p_{z}=0$ \ and integrating the
Eq.(\ref{B11}) we obtain the relation \ $G\,\gamma = G_{0}$ ,
where \ $G=[ 1+R_{0}^{2}( n/n_{0})^{2/3}]^{1/2}$ \ and \ $\gamma
=[ 1 + A^{2}/G^{2}]^{1/2}$ . The straightforward algebra gives the
following relations:

\begin{equation}
n=n_{0}\left( 1-A^{2}/R_{0}^{2}\right )^{3/2} \
\label{B14}
\end{equation}
and
\begin{equation}
G=G_{0}\left[ 1-A^{2}/(1+R_{0}^{2})\right]^{3/2} \ .
\end{equation}
It follows from the Eq.(\ref{B14}) that our considerations remain
valid provided \ $A\leq R_{0}$ ; the plasma density decreases in
the area of EM field localization and if at certain point of this
area \ $A\rightarrow R_{0} $ \ then the plasma density becomes
zero \ ($n\rightarrow 0$), hence, at that point the cavitation
takes place.

The Eq.(\ref{B13}) reduces to the following ordinary differential
equation:
\begin{equation}
\frac{d^{2}a}{d\eta^{2}}-\lambda a + f(a^{2})a=0 \ ,
\label{B15}
\end{equation}
where the nonlinearity function \ $f(a^{2})$ \ is given by
\begin{equation}
f(a^{2})=1 - \frac{\left( 1-a^{2}\right)^{3/2}}{\left(
1-\epsilon^{2}a^{2}\right)^{1/2}} \ .
\label{B16}
\end{equation}
Here \ $\eta=z(\Omega_{e}/c)$ \ is the dimensionless coordinate, \
$\lambda=1-\omega^{2}/\Omega_{e}^{2}$ , \ $a=A/R_{0}$ \ and \
$\epsilon^{2}=R_{0}^{2}/(1+R_{0}^{2})$. One can see that the
nonlinearity function \ $f$ \ is positive and monotonically
increasing function of \ $a$ . For small intensities of EM field \
($a\ll 1$) \ the nonlinearity function is \ $f= (3-\epsilon
^{2})a^{2}/2$ \ while \ $f\rightarrow 1$ \ for \ $a\rightarrow 1$
. Note that the saturating character of the nonlinearity is
related to the plasma cavitation (see Eq.(\ref{B14})). Since \
$0\leq f\leq 1$ \ the Eq.(\ref{B15}) admits the soliton solutions
for all allowed intensities of EM field \ ($0\leq a^{2}\leq 1$) \
provided that \ $0\leq\lambda\leq Max[f]=1$ \ \cite{bib:Vakhitov}.
The paramerer \ $\lambda $ \ is the nonlinear "frequency shift"
and it has the meaning of the reciprocal of the square of the
characteristic width of the soliton.

\bigskip

Integration of Eq.(\ref{B15}) gives
\begin{equation}
\left( \frac{da}{d\eta}\right)^{2}-\lambda a^{2} + F\left(
a^{2},\epsilon\right) = 0 \ .
\label{B17}
\end{equation}
Here we assumed that an integration constant is zero since for the
soliton solutions \ $a,a_{\eta}\rightarrow 0$ \ for \ $\left\vert
\eta\right \vert \rightarrow \infty $ . The nonlinear function \
$F(a^{2},\epsilon)=\int_{0}^{a^{2}}\,f(\xi )
d\xi=F_{1}(a^{2},\epsilon)-F_{1}(0,\epsilon)$ , where \
$F_{1}(a^{2},\epsilon)=(1/4\epsilon^{4})\sqrt{(
1-\epsilon^{2}\,a^{2}) ( 1-a^{2})}\ [3 + \epsilon^{2}(2a^{2}-5)] -
(3/4\epsilon^{5})\,(\epsilon^{2}-1)^{2} \ln
(2\epsilon^{2}\,\sqrt{1-a^{2}} + 2 \, \sqrt{1-\epsilon^{2}a^{2}})$
. From Eq.(\ref{B17}) one can see that the relation between \
$\lambda$ \ and the soliton amplitude \ $a_{m}$ \ is given by \
$\lambda=F(a_{m}^{2},\epsilon ) /a_{m}^{2}$ .

\section{Results and Summary}

The general solution of Eq.(\ref{B15}) cannot be expressed in
terms of the elementary function except for the ultra-relativistic
degenerate plasma case, \ i.e. for \ $R_{0} \gg 1$ \ \
($\epsilon\rightarrow 1$) . Indeed, in this case \ $f=a^{2}$ \ and
for \ $\lambda_{(\epsilon=1)}=a_{m}^{2}/2$ \ the soliton solution
of Eq.(\ref{B15}) obtains a simple form:
\begin{equation}
a=a_{m}sech\left( \frac{a_{m}}{\sqrt{2}}\,x\right) \ .
\label{B18}
\end{equation}

We would like to emphasize that the soliton solution (\ref{B18})
exists for \ $a_{m}=\left( A_{m}/R_{0}\right)\leq 1$ \
($\lambda_{(\epsilon=1)\text{ }}\leq 0.5$) . Consequently we can
state that in the relativistic degenerate plasma the amplitude of
EM soliton can become relativistically strong -- \ $A_{m}\gg 1$ .
In the region of the soliton localization the e-p plasma density
decreases considerably while for \ $A_{m}\rightarrow R_{0}$ \ the
plasma cavitation takes place.

For the nonrelativistic degenerate plasma \ ($\epsilon \ll 1$)\
the nonlinearity function can be approximated by the following
expression: \ $f=1-(1-a^{2} )^{3/2}$ . The Eq.(\ref{B15}) then has
a soliton solution if the nonlinear frequency shift satisfies the
relation \ $\lambda_{(\epsilon=0)}=F(a_{m}^{2},0)/a_{m}^{2} =
1-(2/5)(1-a_{m}^{2})^{5/3}/a_{m}^{2}$ . Thus, the soliton solution
exists for \quad $\lambda _{(\epsilon=0)}\leq0.6$ . Note that \
$A_{m}\leq R_{0}$ \ ($a_{m}\leq 1$) \ and since \ $R_{0}\ll 1$ \
we conclude that in the nonrelativistic degenerate e-p plasma the
EM field intensity is nonrelativistic \ $A_{m}\ll 1$ . However,
even the field intensity is weak it can give rise to a plasma
cavitation at \ $A_{m}\rightarrow R_{0}$. For the intermediate
case when \ $\epsilon < 1 $ \ ($\neq 0$) the soliton
solution exists in plasma 
provided the nonlinear frequency shift \ $\lambda$ \ satisfies an
inequality \
$\lambda_{(\epsilon=1)}<\lambda<\lambda_{(\epsilon=0)}$ \ and
attains its maximal value \ $\lambda_{\max}$ \
($0.5<\lambda_{\max}<0.6$) \ that corresponds to the cavitation \
($a_{m}=1$).

\bigskip

As we emphasized in the introduction the results obtained in this
paper are valid provided that characteristic time of EM field
oscillations is smaller than the pair annihilation time, i.e.
$\kappa\equiv \Omega_e \,\tau_{ann}/2\pi \gg 1 $ , namely for
annihilation time we can use the relation \cite{bib:rohrlich}
\begin{equation}
\tau_{ann}^{-1}=(\pi r_0^2c)\,Q(G_0)\,n \ ,
\label{B19}
\end{equation}
where $r_0$ is an electron radius and

\begin{equation}
Q(x)=\frac{1}{x(1+x)}\,
\left[ \frac{x^2+4x+1}{\sqrt{x^2-1}}\,ln\left(
x+\sqrt{x^2-1}\right)-(3+x) \right] \ .
\label{B20}
\end{equation}

Estimations of parameter \ $\kappa$ \ for different densities
using eq.-s (\ref{B19}) and (\ref{B20}) as well as the expressions
for $G_0$ and $\Omega_e$ given above yield following: \ (1)
$\kappa \sim 1.4 \cdot 10^3$ \ for \ $n\sim n_c = 5.9\cdot
10^{29}\,cm^{-3}$; \ (2) $\kappa \sim 60$ \ for \ $n=5\cdot
10^{32}\,cm^{-3}$; \ (3) $\kappa \sim 7$ \ for \
$n=10^{35}\,cm^{-3}$; \ (4) $\kappa = 0.9$ \ for \
$n=10^{37}\,cm^{-3}$. \ Hence, for the degenerate plasma densities
up to \ $\sim 10^{35}\,cm^{-3}$ \ the conditions for the localized
high frequency pulse solution existence are easily met.

\bigskip

In conclusion, we have considered the possibility of the
high-frequency EM soliton formation in fully degenerate
electron-positron plasma. Applying fully relativistic,
hydrodynamic approach we have shown that such a plasma supports an
existence of stationary soliton solution in over-dense plasma \
($\omega<\Omega_{e}$). In relativistic degenerate e-p plasma the
intensity of EM field can be relativistically strong while for
norelativistic degeneracy case the soliton intensity is always
nonrelativistic. It is also shown that the cavitation of plasma
can occur in both the relativistic and nonrelativistic degenerate
plasmas. The generalization for the case of moving soliton is
straightforward and is beyond of intended scope of present paper.

We believe that the 1-dimensional model of present study can be
generalized for the 2D and 3D problems similarly in either so
called "pancake" regime of propagation with $L_{||}\ll L_{\perp}$
\cite{bib:BM-PRL,bib:fedele,bib:Mora} or in so-called beam-regime
of propagation $L_{\perp}\ll L_{||}$ (where $L_{||}$ and
$L_{\perp}$ are the characteristic longitudinal and transverse
spatial dimensions of the field, respectively) \cite{bib:Ohashi}.
Preliminary analysis shows, that in such cases nonlinear
Shr\"odinger equation with saturating nonlinearity similar to
(\ref{B16}) can be derived implying that the generation of stable
multi-dimensional localized solutions -- "the light bullets" or
the solitary filaments -- is possible.

The results found in given manuscript can be useful to understand
the dynamics of x-ray pulses emanating from the compact
astrophysical objects as well as to study the nonlinear
interactions of intense laser pulses and dense degenerate plasmas
that are relevant for the next-generation intense laser -- solid
density plasma experiments.

\bigskip

N.L. Tsintsadze would like to acknowledge the partial support for
his work from GNSF grant project No. FR/101/6-140/13.



\end{document}